\documentclass[trackchange]{aastex701}

\begin{document}

\title{Inclination of Miranda as a Constraint on the Tidal Dissipation of Uranus}

\author[orcid=0000-0002-4416-8011]{Maryame El Moutamid}
\affiliation{Southwest Research Institute}
\email[show]{maryame.elmoutamid@swri.edu}  

\author[orcid=0000-0003-1226-7960]{Matija \'Cuk}
\affiliation{SETI Institute}
\email{mcuk@seti.org}

\begin{abstract}
The five major moons of Uranus record a complicated geological history characterized by both tectonic resurfacing and heavy impact cratering, connecting their evolution to dynamical processes in the Solar System. Miranda, the smallest and closest major moon to Uranus, has an unusually large orbital inclination ($4.3^\circ$) and an remarkable surface geology dominated by coronae and tectonic structures. In this paper we explore whether Miranda's inclination and geological history can be explained through past orbital resonances within the Uranian satellite system. Using direct $N$-body integrations with the code \textsc{simpl}, we explore the tidal and dynamical evolution of the five major moons while assuming a fast tidal dissipation of Uranus and varying the internal responses of the satellites. Our simulations confirm that Miranda's inclination can be excited through resonant interactions involving Ariel and Umbriel, even when Miranda mean motion is not involved in that resonance. The Ariel-Umbriel 5:3 mean-motion resonance appears to be the most recent major dynamical event capable of driving Miranda's inclination to its current value. We compute the tidal heating associated with these resonance encounters and show that, for a strongly dissipative Uranus ($Q_U\sim600$), the resonance could cause heat fluxes in Miranda approaching those required to form its coronae. These results could provide new constraints on Uranus's tidal dissipation.
\end{abstract}


\section{Introduction}

The moons of the outer planets provide important records of the dynamical and thermal evolution of planetary systems. In particular, tidal interactions and orbital resonances among moons can cause high internal heating and drive long-term orbital evolution. The classical Uranian satellite system, consisting of Miranda, Ariel, Umbriel, Titania, and Oberon, represents a rare example because the moons exhibit both evidence for past geological activity and orbital configuration that lacks current resonances but likely reflects a complicated dynamical history of past resonant encounters.

Miranda is particularly unusual, its $4.3^\circ$ orbital inclination requires substantial dynamical excitation after formation, while its tectonically modified surface indicates episodes of significant internal heating. These two properties may record a common history of past resonant interactions.

Past investigations proposed that Miranda's inclination could be explained through capture into the Miranda-Umbriel 3:1 mean-motion resonance followed by escape after high inclination increase \citep{Tittemore1989,Malhotra1990}. However, subsequent work has shown that the 5:3 mean-motion resonance between Ariel and Umbriel, followed by several secular resonances, can incrementally excite Miranda's inclination \citep{Cuk2020}.

In general, the effects of resonances depends strongly on the tidal dissipation within Uranus. The planetary tidal dissipation factor $Q_U$ governs the rate at which moons migrate (inward or outward) due to tides raised on the planet itself. If Uranus is strongly dissipative ($Q_U \sim 600$), as suggested by  \citep{Jacobson2025}, the moons would migrate much faster and come across resonances more frequently. Conversely, if the planet has a much larger tidal dissipation factor ($Q_U \sim 1.5-2.0\times10^4$), as suggested by past work \citep{Tittemore1990, Cuk2020}, the migration would be much slower and the resonance encounters would be substantially different. 

Because Miranda's inclination is unlikely to have damped significantly over Solar System timescales once excited \citep{Chen2014}, the present orbit of Miranda provides valuable information about the past dynamical evolution of the Uranian satellite system. In this paper we investigate how the coupled tidal and resonant evolution of the Uranian satellites can reproduce Miranda's inclination and what this implies for Uranus's tidal dissipation.

To place this problem in context, we first summarize the dynamical and geological constraints on Miranda’s past evolution.

\section{Background}

The orbital and geological evolution of Uranus's major satellites remains one of the least understood dynamical problems in the outer Solar System. The five classical satellites (Miranda, Ariel, Umbriel, Titania, and Oberon) orbit between approximately $6$ and $23\,R_U$ and likely formed within a circumplanetary disk produced either through standard satellite accretion or through reaccretion following a giant impact associated with Uranus's extreme obliquity \citep{Morbidelli2012,Rufu2022}. Their present-day orbits are nearly circular and coplanar, suggesting that collisional damping and tidal evolution played major roles during their early history. However, the current orbital architecture and geological features of these moons imply that their dynamical evolution was far from inactive.

Among Uranus's satellites, Miranda is the most dynamically anomalous. Its orbital inclination of $4.3^\circ$ is significantly larger than expected for a regular satellite formed in a dissipative circumplanetary disk, where inclinations should be damped to nearly zero. This large inclination therefore requires excitation after formation through some later dynamical event. 
Miranda is also geologically unusual. Voyager 2 observations revealed a surface characterized by massive fault systems, coronae, and resurfaced terrains, including extensional canyons nearly 20~km deep \citep{Thomas1988,Beddingfield2015}. Early geological analyses already suggested that the formation of Miranda's tectonic features required substantial internal heating. In particular, \citet{Pappalardo1997} inferred lithospheric temperature gradients of approximately $8$-$20$~K~km$^{-1}$, corresponding to heat fluxes of roughly $30$-$80$~mW~m$^{-2}$. More recent geological modeling by \citet{Beddingfield2015} estimated heat fluxes of approximately $31$-$112$~mW~m$^{-2}$ for the formation of Arden Corona. These independent geological estimates indicate heat fluxes substantially greater than expected from radiogenic heating alone, supporting an important role for tidal heating in Miranda's geological evolution \citep{Pappalardo1997,Beddingfield2015,Nimmo2023}. These geological constraints naturally raise the question of whether the required heating accompanied the dynamical event that excited Miranda’s orbit.

Early dynamical studies already recognized that the orbital and geological evolution of the Uranian satellites may be closely connected. \citet{Dermott1988} showed that resonant interactions among the Uranian satellites can lead to chaotic orbital evolution and proposed that such dynamical interactions and the associated tidal heating could help explain Miranda's unusual geological activity. This work established an early connection between the complex resonant evolution of the satellite system and Miranda's geological history.

Historically, Miranda's inclination has been attributed to past capture into the Miranda-Umbriel 3:1 mean-motion resonance (MMR). Early studies showed that convergent tidal migration between Miranda and Umbriel could capture Miranda in an inclination-type sub-resonance, causing its inclination to grow until the resonance was broken through secondary resonances or chaotic diffusion \citep{Tittemore1989,Malhotra1989,Malhotra1990}. This framework became the standard explanation for Miranda's present orbit. In this picture, the Miranda-Umbriel 3:1 resonance represented the most recent major dynamical event experienced by Miranda.

More recent numerical work has substantially extended this earlier picture of chaotic orbital evolution. Building on the recognition that resonant interactions can produce chaotic behavior in the Uranian satellite system \citep{Dermott1988}, \citet{Cuk2020} demonstrated that the later Ariel-Umbriel 5:3 resonance can generate chaotic interactions and embedded three-body resonances that perturb all five major satellites. Because Miranda is substantially less massive than the other classical satellites, it is particularly susceptible to these perturbations. Their simulations showed that the Ariel-Umbriel 5:3 resonance alone can excite Miranda's inclination to values comparable to the present-day value, even when Miranda begins with nearly zero inclination. This provides an alternative to the classical interpretation in which Miranda's present inclination is produced primarily by the earlier Miranda-Umbriel 3:1 resonance.

This result significantly alters the interpretation of Miranda's orbital history. If the Ariel-Umbriel 5:3 resonance is sufficient to explain Miranda's present inclination, then the earlier Miranda-Umbriel 3:1 resonance may not be required. Determining whether the 3:1 resonance occurred at all is critical because it directly constrains the total amount of tidal migration experienced by the satellites and therefore the tidal response of Uranus.

The magnitude of tidal migration depends on Uranus's tidal dissipation efficiency, typically parameterized by the tidal dissipation factor $Q_U$ (or more precisely $Q_U/k_{2,U}$), where $k_{2,U}$, the second-degree potential Love number, is a dimensionless parameter that characterizes Uranus elastic response to tidal forcing. It represents the ratio between the gravitational potential generated by the body's tidal deformation and the external perturbing potential, and therefore serves as an important diagnostic of the internal mass distribution and degree of central condensation of the object. The parameter $Q_U/k_{2,U}$ remains highly uncertain. \citet{Cuk2020} inferred relatively weak dissipation, corresponding to $Q_U \sim (1.5-2)\times10^4$, implying modest orbital migration over Solar System history. In contrast, \citet{Nimmo2023} argued that geological constraints from Miranda and Ariel require significantly larger tidal heating rates than can be produced under such weak dissipation, suggesting substantially faster migration. More recently, \citet{Jacobson2025} used astrometric observations to infer a much lower value of $Q_U \sim 600$, implying migration rates nearly two orders of magnitude faster than previously assumed.

These competing estimates lead to fundamentally different evolutionary histories for the Uranian satellites. In low-dissipation scenarios, only a limited number of resonances may have occurred over Solar System history. In high-dissipation scenarios, the satellites may have crossed multiple resonances, including Miranda-Umbriel 3:1, Miranda-Ariel 5:3, Ariel-Titania 4:1, Ariel-Umbriel 5:3, and possibly Ariel-Umbriel 2:1 \citep{Nimmo2023}. Each resonance has distinct dynamical consequences for Miranda's inclination and thermal history.

Taken together, these studies suggest that Miranda preserves a record of both the dynamical and thermal evolution of the Uranian system. Its present inclination constrains which resonances occurred in the past, while its geology constrains the magnitude and timing of tidal heating. By combining direct numerical integrations with estimates of tidal heating during resonance crossings, we aim to determine whether Miranda's present orbit and geology can be explained by a unified dynamical framework and to place new constraints on Uranus's tidal dissipation. We therefore use direct numerical integrations to test how these competing resonance histories depend on the assumed tidal response of Uranus and its satellites. 

\section{Numerical Methods and Results}

\begin{figure}[ht!]
    \centering
    \includegraphics[width=0.9\textwidth]{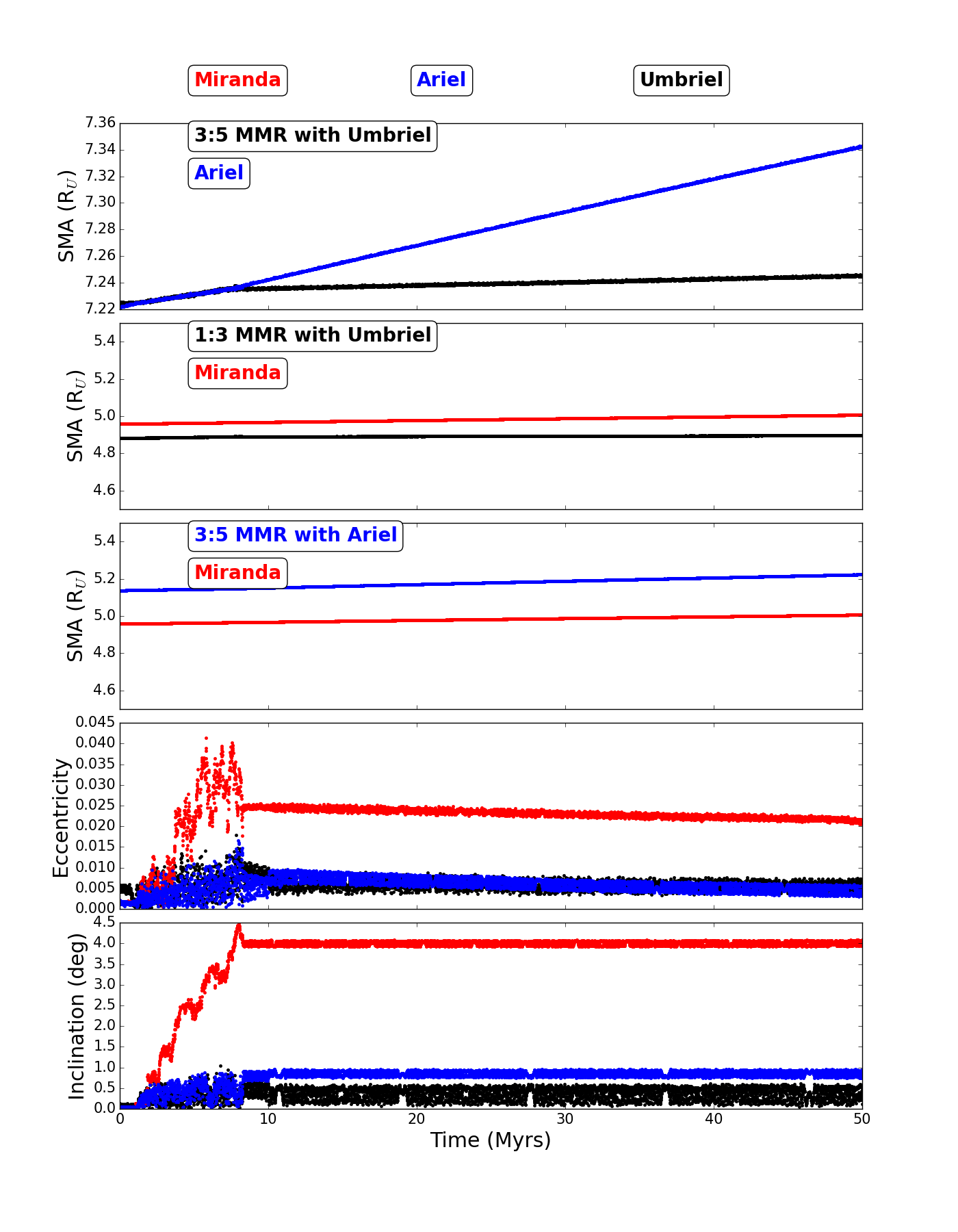}
    \caption{Representative \textsc{simpl} integration for $(k_{2,M},k_{2,A})=(0.1,0.1)$ and $Q_U=600$, following the strong-dissipation estimate of \citet{Jacobson2025}. The simulation shows the coupled orbital evolution of the Uranian satellites. The Miranda-Ariel 5:3 resonance is not encountered during the migration evolution. Instead, the evolution is dominated by the Ariel-Umbriel 5:3 MMR, during which the eccentricities and inclinations of the satellites are strongly excited. In particular, Miranda's eccentricity reaches $\sim0.04$ and its inclination is excited to values comparable to its present value, despite Miranda not being a member of the primary Ariel-Umbriel resonance.}
    \label{fig_aei_k2m01_k2a01}
\end{figure}

\begin{figure}[ht!]
    \centering
    \includegraphics[width=0.9\textwidth]{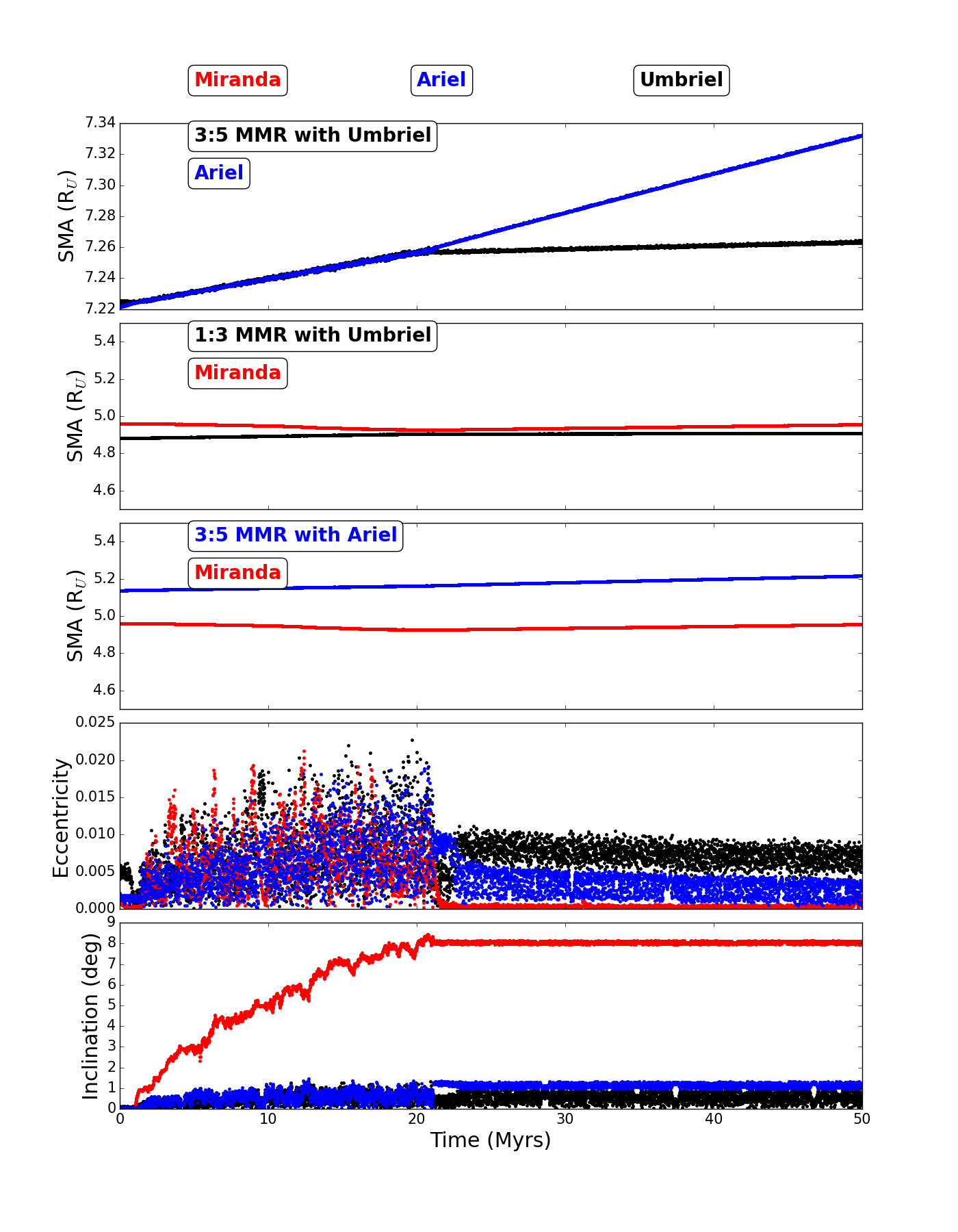}
    \caption{Representative \textsc{simpl} integration for $(k_{2,M},k_{2,A})=(1.0,0.1)$ with $Q_U=600$. The coupled dynamical evolution again avoids a Miranda-Ariel 5:3 resonance encounter. Instead, Umbriel becomes involved in both the Miranda-Umbriel 3:1 and Ariel-Umbriel 5:3 resonances, with the latter interaction governing the long-term evolution of the system.}
    \label{fig_aei_k2m1_k2a01}
\end{figure}

\begin{figure}[ht!]
    \centering
    \includegraphics[width=0.9\textwidth]{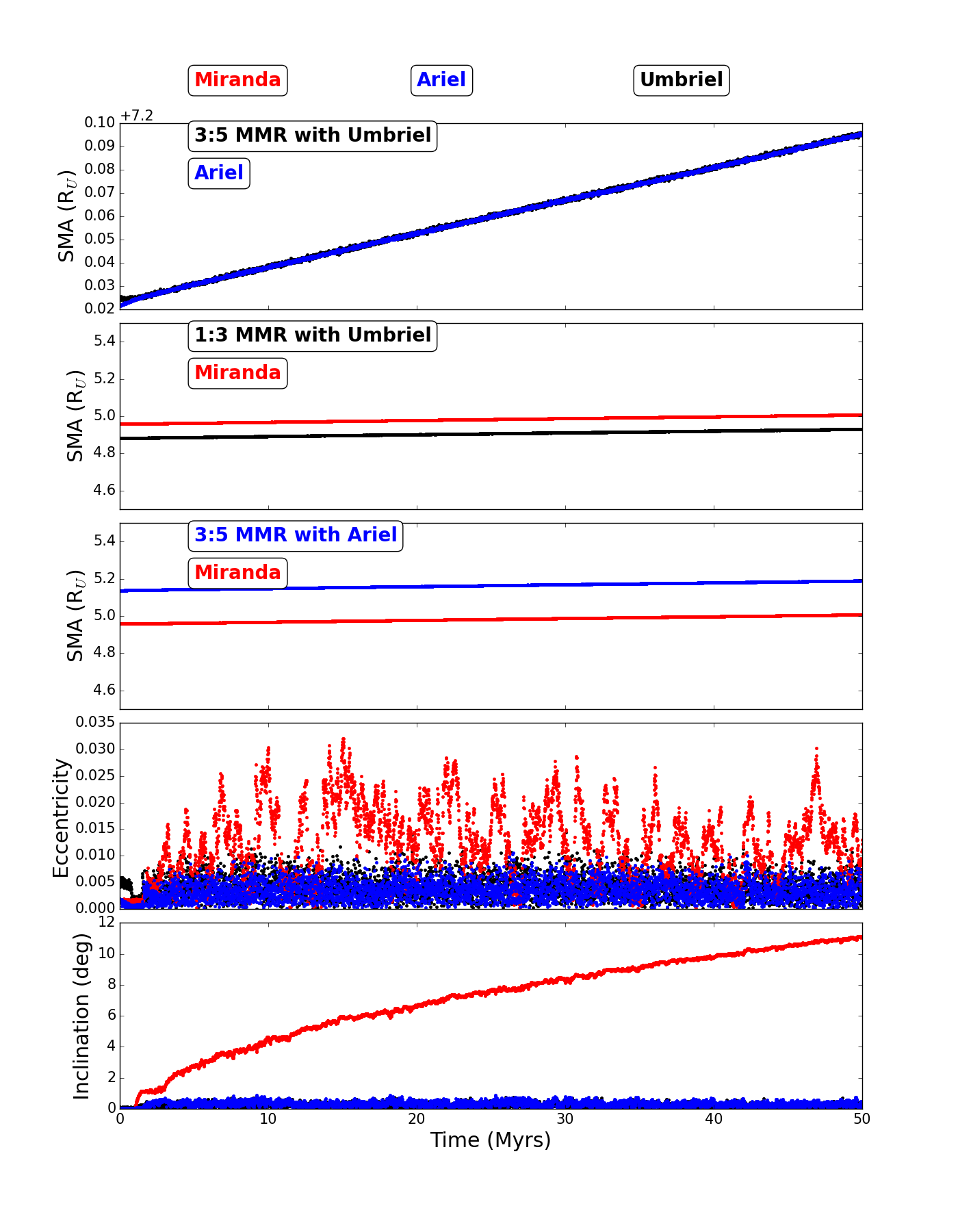}
    \caption{Representative \textsc{simpl} integration for $(k_{2,M},k_{2,A})=(0.1,1.0)$ assuming $Q_U=600$. The migration does not produce a Miranda-Ariel 5:3 resonance crossing. Instead, the evolution is dominated by the Ariel-Umbriel 5:3 MMR. In this simulation, Ariel and Umbriel remain resonantly coupled throughout the $50$~Myr interval shown, and Miranda's inclination consequently continues to increase beyond its present value. This case therefore illustrates the strong indirect excitation of Miranda during prolonged Ariel-Umbriel resonant evolution, but does not represent a complete evolutionary pathway to the present non-resonant configuration.
}
    \label{fig_aei_k2m01_k2a1}
\end{figure}

\begin{figure}[ht!]
    \centering
    \includegraphics[width=0.9\textwidth]{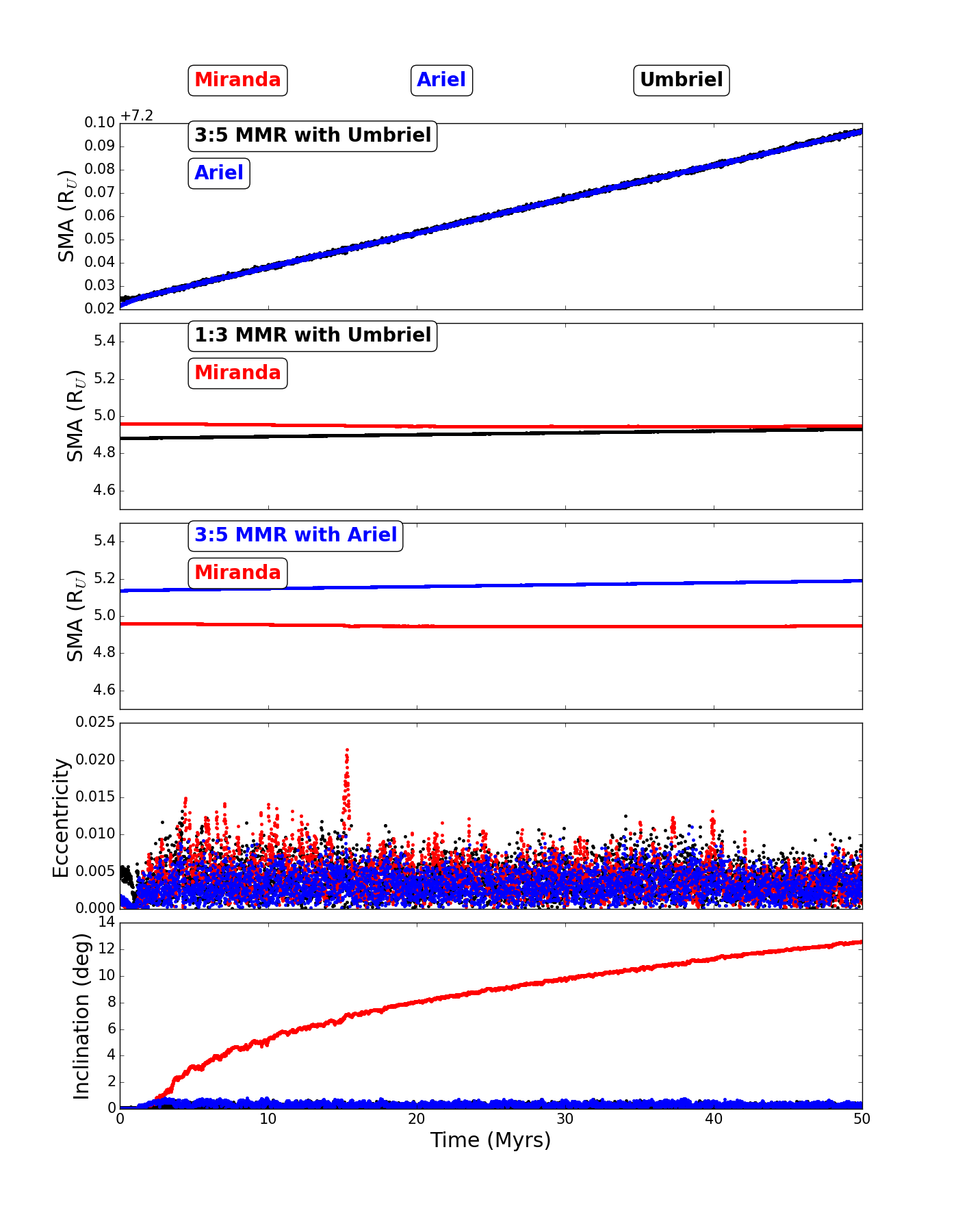}
    \caption{Representative \textsc{simpl} integration for $(k_{2,M},k_{2,A})=(1.0,1.0)$ with $Q_U=600$. The Miranda-Ariel 5:3 resonance is not encountered, while the evolution is dominated by the Ariel-Umbriel 5:3 MMR. Ariel and Umbriel remain resonantly coupled throughout the $50$~Myr interval shown, during which Miranda's inclination continues to increase substantially beyond its present value. Because Ariel and Umbriel are not currently in this resonance, subsequent escape from the 5:3 MMR is required for this case to evolve toward the present satellite configuration.}
    \label{fig_aei_k2m1_k2a1}
\end{figure}

To investigate the dynamical evolution of the Uranian satellites, we performed numerical integrations using the code \textsc{simpl}. This code integrates the gravitational N-body problem while including tidal forces that reproduce the secular effects of equilibrium tides \citep{Cuk2016}.
The simulations include gravitational interactions among Uranus and the five major satellites, as well as the effects of planetary oblateness and the Sun. Tidal forces raised on Uranus drive outward migration of the satellites, while tidal forces raised on the satellites damp orbital eccentricities.
The strength of the planetary tide is controlled by the ratio $k_{2,U}/Q_U$. In the simulations presented here, we adopt $k_{2,U}=0.1$ and explore different values of $Q_U$, with the simulations shown in Figures~\ref{fig_aei_k2m01_k2a01}- \ref{fig_aei_k2m1_k2a1} adopting $Q_U=600$. Thus, the nominal low-$Q_U$ simulations correspond to $k_{2,U}/Q_U=1.67\times10^{-4}$ (or equivalently $Q_U/k_{2,U}=6000$).

For the satellites, we adopt a tidal quality factor $Q=200$ throughout
the integrations. In particular, $Q_M=200$ for Miranda in all four cases
considered below.

\begin{figure}
\centering
\includegraphics[width=0.8\columnwidth]{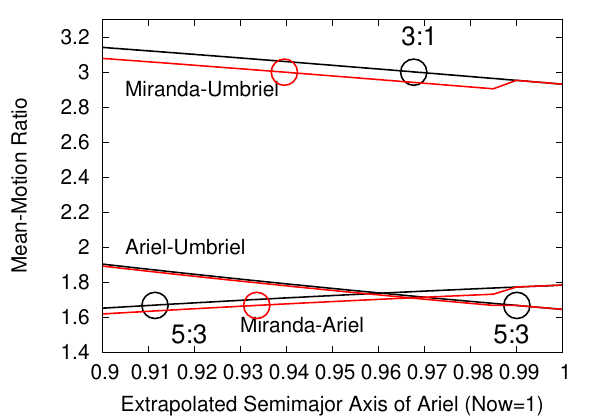}
\caption{
Past two-body mean-motion resonances potentially encountered by the major Uranian satellites during tidal migration. Rather than using absolute time, the horizontal axis is the extrapolated semimajor axis of Ariel normalized to its present value, with $a_A=1$ corresponding to Ariel's current orbit. Because Ariel is the fastest-migrating of the major moons, this parameter provides a convenient proxy for the system's dynamical age. The black curves trace the evolution of the Miranda-Umbriel and Ariel-Umbriel period ratios as the system is evolved backward through progressively smaller values of $a_A$. Circles mark the locations of important commensurabilities, including the Miranda-Umbriel 3:1 and Ariel-Umbriel 5:3 resonances. The red trajectories illustrate the dynamical consequences of the Ariel-Umbriel 5:3 resonance, including temporary parallel migration of Ariel and Umbriel and the inward drift of Miranda driven by tidal coupling among the satellites. This figure emphasizes that the Ariel-Umbriel 5:3 resonance is not merely one resonance among many, but the most consequential recent dynamical event in the system, capable of reorganizing the orbital architecture and indirectly exciting Miranda's orbit.}
\label{fig_res}
\end{figure}

Rather than plotting the evolution directly as a function of time, we parameterize the system evolution using the normalized semimajor axis of Ariel. Because Ariel should be the fastest migrating satellite in the equilibrium tide model, its orbital expansion provides a proxy for time in the resonance evolution of the system. 
Figure~\ref{fig_res} illustrates the calculated past mean-motion resonance encounters among the Uranian satellites as Ariel migrates outward. The horizontal axis shows the normalized semimajor axis of Ariel, while the vertical axis shows the period ratios between Miranda, Ariel, and Umbriel.
Several resonances appear in the expected migration sequence, including the Miranda-Umbriel 3:1 resonance and the Ariel-Umbriel 5:3 resonance. The latter resonance emerges as the most recent significant dynamical transition in the system.

We explored a range of satellite tidal responses by varying the Love numbers of Miranda and Ariel. Four representative cases were considered, corresponding to combinations of $k_{2,M}$ and $k_{2,A}$ equal to $0.1$ or $1.0$. We adopt $Q_M=200$ for Miranda in all four cases; consequently, changing $k_{2,M}$ from $0.1$ to $1.0$ changes $k_{2,M}/Q_M$ from $5\times10^{-4}$ to $5\times10^{-3}$. Because the strength of tidal dissipation within a satellite is controlled by the ratio $k_2/Q$, rather than by $k_2$ alone, the different Miranda Love-number cases therefore correspond to different effective tidal responses.

The four integrations reveal a common dynamical pattern. The Miranda–Ariel 5:3 resonance is not encountered, whereas the Ariel–Umbriel 5:3 resonance occurs in every case and dominates the subsequent evolution. We therefore focus below on how the assumed satellite tidal properties modify Miranda’s response to this resonance.

An important difference among the simulations is the duration of the Ariel-Umbriel 5:3 resonance. In cases~(3) and~(4) (Figures~\ref{fig_aei_k2m01_k2a1} and~\ref{fig_aei_k2m1_k2a1}), Ariel and Umbriel remain resonantly coupled throughout the $50$~Myr interval shown. As a consequence, Miranda's inclination continues to increase and substantially exceeds its present value of $4.3^\circ$. These integrations therefore do not represent complete evolutionary pathways to the present Uranian satellite configuration. Because Ariel and Umbriel are not currently in the 5:3 MMR, an additional mechanism must eventually terminate the resonance in these cases. Possible escape may result from the chaotic evolution within the resonance, interactions with additional resonant or changes in the tidal migration rates. Determining the specific escape mechanism and its timescale is beyond the scope of the present integrations. Nevertheless, these cases demonstrate that prolonged capture in the Ariel-Umbriel 5:3 resonance would over-excite Miranda's inclination, providing an additional constraint on the allowed duration of this resonance in the past.

However, these apparently over-excited cases cannot be ruled out solely
from Miranda's present inclination, because subsequent tidal damping
could reduce the inclination, particularly if Miranda possessed a
subsurface ocean during part of its thermal evolution
\citep{Chen2014}. The resonance-driven eccentricity excitation also has important thermal consequences for Miranda.

To quantify the thermal consequences of the resonance-driven eccentricity
excitation, we calculate the instantaneous tidal heating of Miranda using
the standard expression for eccentricity tides in a synchronously rotating
satellite,
\begin{equation}
\dot{E}_{\rm tide}
=
\frac{21}{2}
\frac{n^5 R_M^5}{G}
\frac{k_{2,M}}{Q_M}
e_M^2,
\label{eq:tidalheating}
\end{equation}
where $n$ and $e_M$ are Miranda's instantaneous mean motion and
eccentricity obtained directly from each simulation, $R_M=235$~km is
Miranda's radius, and $Q_M=200$. The corresponding globally averaged
surface heat flux is
\begin{equation}
F_{\rm tide}
=
\frac{\dot{E}_{\rm tide}}{4\pi R_M^2}.
\end{equation}
Because geological structures are expected to preferentially record
episodes of strongest tidal activity, we report the peak tidal
power and peak surface heat flux reached during the relevant
Ariel-Umbriel 5:3 resonance episode, rather than a time-averaged heating
rate. For each simulation, the peak value is evaluated using the maximum
eccentricity attained by Miranda during that resonance encounter. Applying this expression to the maximum eccentricity reached during the 5:3 episode yields the peak heating rates summarized in Table 1.

The peak heating estimates show that several of the simulated
tidal-response cases are capable of producing geologically significant
heating of Miranda. In case~(1), Miranda reaches
$e_M\sim0.04$, yielding a peak tidal power of
$\sim29{,}000$~MW and an averaged peak surface heat flux of
$\sim42$~mW~m$^{-2}$. This lies within the
$\sim31$-$112$~mW~m$^{-2}$ range estimated by
\citet{Beddingfield2015} for the formation of Arden Corona.

Cases~(2) and~(4), which adopt $k_{2,M}=1.0$, reach lower maximum
eccentricities of approximately $0.02$, but their larger
$k_{2,M}/Q_M$ produces substantially greater peak tidal dissipation.
Both cases reach peak tidal powers of approximately
$7.2\times10^4$~MW, corresponding to surface heat fluxes of
$\sim104$~mW~m$^{-2}$, also within the range inferred for Arden Corona.
Case~(3), with $k_{2,M}=0.1$ and a maximum eccentricity of
$\sim0.03$, produces a lower peak heat flux of approximately $24$~mW~m$^{-2}$.

These results illustrate that the thermal response depends jointly on the
resonance-driven eccentricity excitation and the assumed satellite tidal
parameter $k_{2,M}/Q_M$. A simulation producing a smaller peak
eccentricity can therefore generate greater tidal heating if the adopted $k_{2,M}/Q_M$ is sufficiently large.

Figures~\ref{fig_aei_k2m01_k2a01},~\ref{fig_aei_k2m1_k2a01},~\ref{fig_aei_k2m01_k2a1} and \ref{fig_aei_k2m1_k2a1} show the four representative integrations. Although all cases undergo the Ariel–Umbriel 5:3 resonance, Miranda’s eccentricity and inclination response depends strongly on the assumed tidal properties of Miranda and Ariel. Case (1) is particularly noteworthy because it reproduces Miranda’s present inclination without requiring long-lived capture into the earlier Miranda–Umbriel 3:1 resonance.

\section{Miranda's inclination as a record of Uranus's dynamical history}

The simulations above change how Miranda’s present inclination can be interpreted as a record of the system’s resonance history. In particular, excitation during the Ariel–Umbriel 5:3 resonance reduces the need to invoke the earlier Miranda–Umbriel 3:1 resonance as the primary source of Miranda’s present inclination.

We find that Miranda's inclination can be reproduced through indirect excitation during the later Ariel-Umbriel 5:3 resonance, even when Miranda begins with nearly zero inclination. During this resonance crossing, chaotic interactions and embedded three-body resonances perturb all major satellites, but Miranda experiences the strongest response because it is the least massive satellite in the classical Uranian system. This confirms and extends the results of \citet{Cuk2020}.

More importantly, our simulations indicate that Miranda's inclination behaves differently from other orbital elements during chaotic resonance evolution. While eccentricities and inclinations of the larger satellites tend to fluctuate stochastically around quasi-equilibrium values, Miranda's inclination often exhibits cumulative long-term increasing. Even in simulations where Miranda begins with its present-day inclination, the Ariel-Umbriel 5:3 resonance frequently drives additional inclination increasing. 

This behavior has important implications for earlier resonance crossings. If Miranda previously experienced strong excitation during the Miranda-Umbriel 3:1 resonance, its inclination may have already been large before the later Ariel-Umbriel 5:3 resonance occurred. Because the later resonance appears to further increase Miranda's inclination, this could produce final inclinations substantially larger than the currently observed value.

This apparent over-excitation admits two possible interpretations: either Miranda avoided substantial excitation during the earlier Miranda–Umbriel 3:1 resonance, or its inclination was subsequently reduced by tidal damping.

Alternatively, Miranda’s inclination may have been partially damped after resonant excitation. Tidal inclination damping can be substantially enhanced in satellites with subsurface oceans \citep{Chen2014}, and independent dynamical and thermal arguments suggest that oceans may have existed within the Uranian satellite system \citep{Cuk2020,Bierson2022}. Miranda’s present inclination therefore need not represent its maximum past excitation. If ocean-enhanced damping operated efficiently, stronger resonance excitation, including excitation during an earlier Miranda–Umbriel 3:1 encounter, cannot be excluded. The present inclination therefore constrains the coupled orbital and thermal evolution of Miranda, rather than the dynamical history alone.

An additional complication is that the relationship between $Q_U$ and the total amount of past orbital evolution is not monotonic. During and after the Ariel–Umbriel 5:3 resonance, Miranda can migrate inward (Fig.~\ref{fig_aei_k2m01_k2a01}), partially offsetting earlier outward evolution. Even modest inward migration shifts the locations and timing of previous resonance encounters, particularly the Miranda–Umbriel 3:1 MMR. Consequently, the present satellite configuration cannot be mapped uniquely onto a single tidal-evolution timescale or value of $Q_U$.

The prolonged captures in cases (3) and (4) further show that Miranda’s present inclination constrains the duration, not merely the occurrence, of the Ariel-Umbriel 5:3 resonance.

These dynamical constraints can now be considered together with recent estimates of Uranus’s tidal dissipation.

\section{Discussion and Conclusion}

The dynamical and thermal results above place Miranda in a broader context: its present inclination constrains not only the occurrence of individual resonances, but also the cumulative tidal evolution of the Uranian satellite system. 

Our results can be interpreted in the context of recent studies addressing Uranus's tidal evolution. \citet{Cuk2020} argued that Uranus likely possesses a relatively high tidal dissipation factor, $Q_U \sim 1.5\times10^4$-$2\times10^4$, corresponding to weak dissipation and slow orbital migration of the major satellites. In such a scenario, the resonance history of the system evolves over long timescales, and Miranda's present inclination reflects the cumulative effects of past resonant interactions. Our simulations remain broadly consistent with this work in the sense that the Ariel-Umbriel 5:3 resonance still emerges as a key event even for high-$Q_U$ histories. However, the tidal heating generated in these weak-dissipation cases remains relatively modest. If Miranda's coronae formed in response to resonance-driven tidal heating, then an additional amplification mechanism may be required, such as localized dissipation within a partially molten interior or an earlier epoch of stronger dissipation.

In contrast, \citet{Nimmo2023} proposed that the Uranian satellites may preserve evidence for much stronger dissipation than traditionally assumed. His work opened the possibility that Uranus's internal structure permits considerably faster tidal evolution. Our simulations support this interpretation. In low-$Q_U$ cases, resonance crossings become far more dynamically consequential, and the Ariel-Umbriel 5:3 resonance can generate enough tidal heating within Miranda to drive major geological activity. Miranda therefore provides a direct link between dynamical and geophysical evolution: the same resonance capable of reproducing the present orbit may also explain the moon's resurfacing history.

At the same time, our simulations suggest an important limitation on extremely strong dissipation scenarios. Faster tidal migration implies that the satellites would have crossed a larger number of resonances over Solar System history, including not only the Ariel-Umbriel 5:3 resonance but also earlier resonances such as Miranda-Umbriel 3:1, Miranda-Ariel 5:3, Ariel-Titania 4:1, and Ariel-Umbriel 2:1 \citep{Nimmo2023}. Because Miranda's inclination tends to increase cumulatively during chaotic resonance evolution, repeated strong resonance passages can over-excite the orbit and produce inclinations much larger than the currently observed value of $4.3^\circ$. In this sense, Miranda's present inclination may place an upper limit on the total number and strength of past resonance crossings.

A third framework was recently proposed by \citet{Jacobson2025}, who derived a present-day estimate of Uranus's tidal dissipation factor near $Q_U\sim600$, implying dissipation nearly thirty times stronger than the values adopted by \citet{Cuk2020}. Such a low-$Q_U$ regime has major consequences for the Uranian system. In our simulations, stronger dissipation produces larger orbital migration, modifies the spacing at which resonances are encountered, and substantially increases the tidal heating generated during resonance passages. Under these conditions, the Ariel-Umbriel 5:3 resonance becomes not only the dominant dynamical event shaping Miranda's orbit but also a plausible trigger for its most recent tectono-thermal activity. 

These conclusions are further supported by recent studies of resonance evolution in fast-migration scenarios. \citet{Rossi2026} revisited the Ariel-Umbriel 2:1 resonance and showed that rapid outward migration makes resonance crossing and capture highly likely. Their work demonstrated that escape from the resonance may require later interactions involving Titania and three-body resonant chains, while also generating substantial tidal heating within Ariel. By contrast, \citet{Gomes2026} investigated the divergent crossing of the Miranda-Ariel 7:4 resonance and found that it produces only limited eccentricity excitation and does not significantly alter the later evolution of the system. Their results suggest that not all resonances contribute equally to the present architecture of the Uranian satellites and further support the idea that the Ariel-Umbriel 5:3 resonance was the dominant recent event governing Miranda's evolution.

Taken together, these comparisons indicate that the question of whether Uranus is best described by a high-$Q_U$ or low-$Q_U$ evolutionary history is not completely resolved. 
But our results suggest that Miranda's present inclination and coronae may both be consequences of a relatively dissipative Uranus combined with small tidal responses of the satellites themselves. 
Miranda's inclination and surface geology may provide the critical observables needed to constraint Uranus dissipation. If Miranda's coronae were indeed produced during the Ariel-Umbriel 5:3 resonance, then either Uranus is dissipative, or its dissipation varied over time. The latter possibility would imply that Uranus's tidal response depends on forcing frequency, interior structure, or episodic resonance locking \citep{Fuller2016} rather than behaving as a simple constant-$Q$ planet. In this sense, Miranda is not only an anomalous moon but a probe of Uranus itself, preserving a record of the planet's long-term dynamical and thermal evolution.


\begin{deluxetable*}{lccccc}
\tablecaption{
Peak tidal heating of Miranda during the Ariel- Umbriel 5:3 mean-motion
resonance for simulations assuming a strongly dissipative Uranus
($Q_U=600$). The reported values correspond to the maximum eccentricity
reached by Miranda during the relevant resonance episode in each simulation.
The tidal power is calculated assuming $Q_M=200$ and using
$\dot{E}_{\rm tide}=(21/2)(n^5R_M^5/G)(k_{2,M}/Q_M)e_M^2$.
Surface heat fluxes are obtained from
$F_{\rm tide}=\dot{E}_{\rm tide}/(4\pi R_M^2)$, adopting
$R_M=235$~km. Geological constraints indicate that substantial heat fluxes were required during Miranda's tectonic evolution. \citet{Pappalardo1997} inferred lithospheric temperature gradients of approximately $8$- $20$~K~km$^{-1}$, corresponding to heat fluxes of roughly $30$- $80$~mW~m$^{-2}$, while \citet{Beddingfield2015} estimated heat fluxes of approximately $31$- $112$~mW~m$^{-2}$ for the formation of Arden Corona.}
\label{tab_TH}
\tablewidth{0pt}
\tablehead{
\colhead{} &
\colhead{\v{C}uk et al. (2020)} &
\colhead{Case (1)} &
\colhead{Case (2)} &
\colhead{Case (3)} &
\colhead{Case (4)}
}
\startdata
$k_{2,M}$
& \nodata
& 0.1
& 1.0
& 0.1
& 1.0 \\
$Q_M$
& \nodata
& 200
& 200
& 200
& 200 \\
$k_{2,M}/Q_M$
& \nodata
& $5\times10^{-4}$
& $5\times10^{-3}$
& $5\times10^{-4}$
& $5\times10^{-3}$ \\
$e_{M,\max}$
& \nodata
& $\sim0.04$
& $\sim0.02$
& $\sim0.03$
& $\sim0.02$ \\
Peak Tidal Heating (MW)
& 200
& 29,000
& 72,500
& 16,300
& 72,500 \\
Peak Surface Heat Flux (mW m$^{-2}$)
& 40$^{*}$
& 42
& 104
& 24
& 104 \\
\enddata

\tablecomments{
Case (1): $(k_{2,M},k_{2,A})=(0.1,0.1)$;
Case (2): $(k_{2,M},k_{2,A})=(1.0,0.1)$;
Case (3): $(k_{2,M},k_{2,A})=(0.1,1.0)$;
Case (4): $(k_{2,M},k_{2,A})=(1.0,1.0)$.
The values reported for Cases (1)- (4) are instantaneous peak values,
evaluated at the maximum Miranda eccentricity reached during the relevant
Ariel- Umbriel 5:3 resonance episode, rather than averages over the
resonance evolution. \\
$^{*}$ Value reported in \citet{Cuk2020} was incorrect. 
}
\end{deluxetable*}

\clearpage
\newpage

\begin{acknowledgments}

This work was supported by the NASA Science Mission Directorate - Solar System Workings Program \#80NSSC24K1842.

\end{acknowledgments}

\bibliographystyle{aasjournal}
\bibliography{bib_refs_main}

\end{document}